\documentclass[aps,floatfix,showpacs,tightenlines,twocolumn,superscriptaddress]{revtex4}
\usepackage{times}
\usepackage{graphicx}
\usepackage{epsfig}
\begin{document}

\title{Teleportation of composite systems for communication and information processing}

\author{Sebastien G.R. Louis}\email{seblouis@nii.ac.jp}
\affiliation{National Institute of Informatics, 2-1-2
Hitotsubashi, Chiyoda-ku, Tokyo 101-8430, Japan}
\affiliation{Department of Informatics, School of Multidisciplinary Sciences,
The Graduate University for Advanced Studies,
2-1-2 Hitotsubashi, Chiyoda-ku, Tokyo 101-8430 Japan}

\author{Andrew D. Greentree}
\affiliation{School of Physics, University of Melbourne, Parkville, Victoria 3010, Australia}
\affiliation{Centre for Quantum Computer Technology, School of Physics,
University of Melbourne, Parkville, Victoria 3010, Australia}

\author{W. J. Munro}
\affiliation{Hewlett-Packard Laboratories, Filton Road, Stoke
Gifford, Bristol BS34 8QZ, United Kingdom}
\affiliation{National Institute of Informatics, 2-1-2
Hitotsubashi, Chiyoda-ku, Tokyo 101-8430, Japan}

\author{Kae Nemoto}
\affiliation{National Institute of Informatics, 2-1-2
Hitotsubashi, Chiyoda-ku, Tokyo 101-8430, Japan}

\begin{abstract}
We describe two protocols for efficient data transmission using a single passive bus. Different types of interactions are obtained enabling  deterministic transfer and teleportation of composite quantum systems for arbitrary subsystem dimension and for arbitrary numbers of subsystems. The subsystems may become entangled in the transmission in which case the protocols can serve generalized teleportation based information processing as well as storage and transmission functions. We explore the cases of two qubits and two qutrits in detail, obtaining a maximally entangling mapping of the composite systems and discuss the use of a continuous variable bus.
\end{abstract}


\maketitle

\section{Introduction}  

In recent years there has been a tremendous amount of research done in harnessing the properties of quantum objects to perform information processing \cite{niel-chuang}. This is due in part to the potential speed-up of quantum algorithms over classical algorithms for some computational tasks \cite{shor,ekert,grover} and the possibility to directly simulate quantum systems. To achieve these tasks requires a scalable quantum information processor. On the fundamental level this means choosing a physical qubit realization and the interaction which enables the implementation of quantum logical gates. On a higher level we need to organize the constituents such as qubits, measuring devices or gates, within a finite physical space in a scalable fashion, including both transport and concatenated error correction for strong scalability \cite{hol}. 

Many of the proposed quantum computer architectures include spatially distinct regions that perform the roles of memory and interaction \cite{thaker,rod,kielp,tayl,hol}. Such an approach presents several advantages, the first being the suppression of decoherence in well-isolated memories. Another key feature of this type of architecture is the extendibility of the system. New areas of memory can be added, physically keeping the same processing area but reprogramming it accordingly. This is a vital attribute when we consider large scale applications. There is also potential for simplifying and concentrating the level of control needed, and mitigating the effects of cross-talk, by restricting the number of control elements in the processing regions. Finally, a level of defect-tolerance can be incorporated by routing around defective regions. 

In this context an efficient transfer of information from the memory areas to the processing areas is crucial \cite{div}. To achieve this information transfer, current proposals include the use of mobile qubits \cite{hol,kielp,tayl,skinner} and flying qubits with an interconversion to stationary qubits \cite{duan1,duan2,oi}. Other possible frameworks for data transfer are spin chains \cite{bose,lloyd,fries} and quantum bus schemes \cite{tim,loock-rep}. Teleportation can also be used in quantum computer architectures \cite{gotc,rod-multi} to provide effective communication and computation channels.

Given an interface between stationary and flying systems, one natural question is: how could higher dimensional buses be used in such data transfer schemes? This constitutes the central theme of the present work. For example we might want to transmit a pair of qubits with a single use of a quantum channel. In general, the efficient use of qudits can optimize the Hilbert space of the system's degrees of freedom \cite{andy}. Most of the qubit realizations proposed and used are actually embedded in a qudit structure already with the non-computational states seen as sources of potential error to be quantified and mitigated \cite{dev}.

The study of qudits in information processing and communication has generated many results \cite{bart,jam,leary,xen}, defining generalized gates, teleportation protocols and finding feasible physical implementations \cite{hugh}. Additionally, the transient occupation of higher dimensional states can greatly reduce the complexity of certain gates, for example Ralph \textit{et al.} have shown that the efficiency of synthesising the Toffoli gate can be improved by using a qutrit subspace \cite{ralph}. Yet the issue of data transfer between arbitrary dimensional systems through a single higher dimensional qudit bus has not been considered. Such a qudit bus would constitute a generic resource, enabling the distribution of entanglement and data over different groups of systems in a flexible fashion. This will result in a physical compression of the information, reducing the number of controlled physical systems and the number of quantum channels required across the processor. 


Here we show protocols for high dimensional quantum transfer employing a passive mediating bus. By keeping this mediator passive (fixing it as the target to all qudit gates and avoiding local operation on it), we simplify the interactions and reduce the level of control needed. The information held by an arbitrary composite system can either be transfered or teleported via the bus to a recipient system in another location, through entangling operations, measurements and feed-forward. We focus initially on a composite system made up of two subsystems of equal dimension and then generalize to arbitrary numbers of subsystems. To illustrate our scheme we describe in detail the cases of two-qubit and two-qutrit composite systems. As the composite system is being transmitted, non-trivial operations may also be applied. 


This paper is organized as follows: the protocols and their requirements are introduced in Section II, before we consider in detail the example of two qubits in Section III. Section IV provides two types of interactions insuring a deterministic transfer for systems of arbitrary dimension and we use these methods to explore the case of two qutrits in the Appendix. In section V we propose the use of a continuous variable bus before summarizing the results and pointing toward future work in the conclusions. 

\section{Protocols}

\begin{figure}
\begin{center}\includegraphics[scale=0.45]{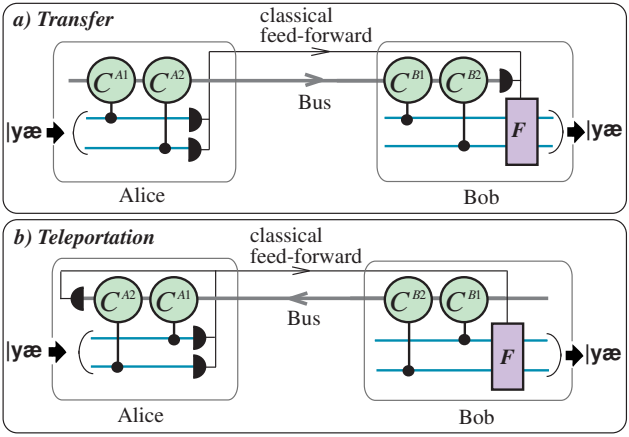}\end{center}
\caption{Schematic representation of the two variants of the protocol. In a) Alice first couples her input state $|\psi\rangle$ with the passive bus via conditional unitary operations and measures out her two subsystems in a conjugate basis. She sends the bus and the measurement results to Bob. On his side, Bob has prepared the recipient state of two subsystems and on receiving the bus, couples his subsystems  to the bus via conditional unitary operations. After measuring the bus, Bob performs feed-forward (denoted by a unitary operation $F$) on his state to reconstruct Alice's input. In b) Bob first couples his recipient state to the bus and sends it to Alice. Upon receiving the bus she couples her input state with it and then proceeds with the measurements as in the transfer protocol. All the results are then communicated to Bob who performs the adapted feed forward, effecting qudit teleportation.}
\end{figure}

Our protocols enable quantum communication between two parties, Alice and Bob, via a passive bus.  We assume initially that Alice has two subsystems (qudits) of equal dimension that she wishes to send to Bob, who also has two qudits of the same dimension as Alice (see Fig. 1). Initially Alice holds two $d$-dimensional systems $A1$ and $A2$ in an arbitrary state $|\psi\rangle_A=\sum_{i,j=0}^{d-1}x_{ij}|i\rangle_{A1}|j\rangle_{A2}$. Initiating the transfer protocol, Alice couples her composite system to the $d^2$-dimensional bus via conditional unitary operations. These can be written as
\begin{equation}
C^{Aj}=\sum_{i=0}^{d-1}|i\rangle\langle i|\otimes U_i^{Aj},
\label{}
\end{equation}
where the projectors $|i\rangle\langle i|$ act on subsystem $Aj$ (here $j=1$, 2) and the unitary operations $U_i$ act on the bus state. An appropriate set of operations for each subsystem will produce a one-to-one mapping between the basis states of the composite system and the basis states of the bus (always to the right, with basis states $|\varphi_{ij}\rangle$), guaranteeing a complete mapping of the $x_{ij}$ coefficients. The resulting combined state we write as
\begin{eqnarray}
|\xi\rangle=\sum_{i,j=0}^{d-1}x_{ij}|i\rangle_{A1}|j\rangle_{A2} |\varphi_{ij}\rangle,\nonumber\\
\quad\mathrm{with}\quad\langle \varphi_{i'j'}|\varphi_{ij}\rangle=\delta_{ii'}\delta_{jj'},
\label{}
\end{eqnarray}
at which point the bus is then sent to Bob through a quantum channel.

Before receiving the bus, Bob prepares his two $d$-dimensional recipient qudits in the equally weighted superposition $|\psi'\rangle_B=\frac{1}{d}\sum_{k,l=0}^{d-1}|k\rangle_{B1}|l\rangle_{B2}$. Then he couples each one of them to the encoded bus via interactions of the form (1), leading to a combined state
\begin{equation}
C|\psi'\rangle |\xi\rangle=\frac{1}{d}\sum_{i,j,k,l=0}^{d-1}x_{ij}(|k\rangle|l\rangle)_B(|i\rangle|j\rangle)_A U_l^{B2} U_k^{B1} |\varphi_{ij}\rangle,
\label{}
\end{equation}
with $C=C^{B2}C^{B1}$. To transfer the input state, Alice measures her subsystems in a conjugate basis (one can be obtained through a Fourier transform of the computational basis). This can be done at any time after sending the bus, removing $|i\rangle|j\rangle$ from the above expression up to known phases. The results will be sent as classical information used in the final feed-forward applied by Bob.

To complete the transfer, Bob measures the mediator and \textsl{for all measurement results} retrieves Alice's state up to a known correction (unitary two-qudit operation, denoted by $F$ in Fig. 1). Complete quantum information transfer places requirements on the unitary operations, $\left\{U_k^{B1},k=0,1,..,d-1\right\}$ and $\left\{U_l^{B2},l=0,1,..,d-1\right\}$ that must be fulfilled. These requirements can be expressed thus
\begin{equation}
\mathrm{Tr}\left[\left(U_l^{B2}U_k^{B1}\right)\left(U_{l'}^{B2} U_{k'}^{B1}\right)^{\dag}\right]=d^2\delta_{kk'}\delta_{ll'},
\label{}
\end{equation}
for all $k,k',l$ and $l'$. The above expression states that any ordered combination made up of a single unitary operation from each set needs to result in an operation orthogonal to all other combinations, in terms of the Hilbert-Schmidt inner product, defined on operators $\hat{V}$ and $\hat{W}$ as $\mathrm{Tr}(\hat{V}\hat{W}^{\dag})$.

Reversing the order of the coupling to the mediator allows qudit quantum teleportation to be performed [Fig. 1 (b)]. In this case, Bob first entangles his subsystems (prepared in an equally weighted superposition, as before) with the mediating bus, and sends the mediator to Alice. Alice then entangles her state with the mediator. The entanglement and subsequent measurement enables the completion of a qudit teleportation protocol between Alice and Bob. Keeping the indices used above, the final state after these interaction is precisely (3), switching $i$ and $j$ for $k$ and $l$. Thus a deterministic transfer of the quantum information held by Alice's composite system is obtained if the unitary operations contained in her interactions obey the relation (4). In other words we have flexibility in the direction in which we want to use the quantum channel, leading to two different protocols, serving essentially the same purpose and requiring the same type of interactions. 

Now we must identify the sets of unitary operations that satisfy (4), and for this we focus on a particular class of unitary operators namely permutation operators. These operators we define as $P\equiv\sum_{s=0}^{m-1}|p(s)\rangle\langle s|$ where $p$ is a  permutation mapping an ordered set of elements to itself, written as $p(s)=s'$. A compact expression for describing permutations is provided by the cycle notation \cite{bauer}
\begin{equation}
\left( \begin{array}{ccccc}
1 & 2 & 3 & 4 & 5\\
2 & 3 & 1 & 5 & 4\end{array} \right)\equiv(123)(45),
\label{}
\end{equation} 
where each pair of brackets contains a cycle which is read from left to right. The effect of $p$ on an element can for example be written as $p(4)=5$. The operator corresponding to (5) is then $P=|1\rangle\langle 3|+|2\rangle\langle 1|+|3\rangle\langle 2|+|4\rangle\langle 5|+|5\rangle\langle 4|$ and the associated permutation $p$ entirely specifies the operator $P$.

Having chosen and defined the class of permutation operators, we proceed to writing down the two sets of operators $\left\{P_k^{B1},k=0,1,..,d-1\right\}$ and $\left\{P_l^{B2},l=0,1,..,d-1\right\}$, for each subsystem. In addition to the orthogonality requirements, by choosing one of the permutations in each set ($P_0^{B1}$ and $P_0^{B2}$) to be the identity, expression (4) implies that all non-trivial combinations must correspond to \textit{complete} permutations (derangements). This can be expressed as $P^{B2}_lP^{B1}_k|s\rangle \neq|s\rangle$ for all $k$ and $l$ except when $k=l=0$. The simplest case occurs for $d$=2, which we explore in detail in the next section.

\section{Transmitting two qubits via one ququad}

To illustrate our transfer protocol we consider the transmission of a two-qubit state. To effect transmission, Alice and Bob require a four dimensional bus, i.e. a ququad. There is a total of $n!$ permutations on $n$ elements, of which $!n=n!\sum_{k=0}^n (-1)^k/k!$ correspond to complete permutations \cite{mont}. In consequence, given the present dimensionality, we have $!4=9$ permutation operators to choose from. We define the bus basis states $\{|s\rangle,s=0,..,3\}$. The full interaction between Alice's two qubits and the bus we write as
\begin{eqnarray}
C^{A2}C^{A1}=\left(|0\rangle_{A2}\langle0|\otimes I^{A2}+|1\rangle_{A2}\langle1|\otimes P^{A2}\right)\nonumber\\
\times\left(|0\rangle_{A1}\langle0|\otimes I^{A1} +|1\rangle_{A1}\langle1|\otimes P^{A1}\right),
\label{}
\end{eqnarray}
where, the identity $I$ and the permutation operators $P^{A1}$ and $P^{A2}$ act on the bus. We will arrange the possible operators into two groups, one consisting of pairwise swap operations and the other of cyclic permutations. They are represented schematically in Fig. 2. There are 3 distinct pairwise swap permutations which in the cycle notation we write as $q_1=(01)(23)$ (corresponding to the permutation operator $Q_1=|1\rangle \langle 0|+|0\rangle\langle 1|+|3\rangle\langle 2|+|2\rangle\langle 3|$), $q_2=(02)(13)$ and $q_3=q_1q_2=(03)(12)$. The 6 cyclic permutations are given by $r_1=(0123)$, $r_2=(0132)$, $r_3=(0213)$ and their inverses. We begin with the first type of interaction in which both Alice and Bob make use of pairwise swap operators. Proceeding with the first part of the transfer protocol, Alice starts with her two qubits in an arbitrary state with the bus initiated in the $|0\rangle$ state, leading to a combined state
\begin{equation}
|\psi\rangle=\left(x_0|00\rangle+x_1|01\rangle+x_2|10\rangle+x_3|11\rangle\right)_A|0\rangle.
\label{}
\end{equation} 
Setting $P^{A1}=Q_1$, $P^{A2}=Q_3$ she entangles her state with the bus,
\begin{equation}
C|\psi\rangle=x_0|00\rangle_A|0\rangle+x_1|01\rangle_A|3\rangle+x_2|10\rangle_A|1\rangle+x_3|11\rangle_A|2\rangle,
\label{}
\end{equation}
with $C=C^{A2}C^{A1}$. Then she measures out her qubits in the $|+/-\rangle$ basis and up to phase corrections depending on the measurement outcomes, Alice sends the disentangled bus to Bob, which is in state
\begin{equation}
|\xi\rangle=x_0|0\rangle+x_1|3\rangle+x_2|1\rangle+x_3|2\rangle.
\label{}
\end{equation}
The phase corrections are sent as classical information and kept until then end of protocol when Bob performs the feed-forward operation on his two-qubit state.  
\begin{figure}[!htb]
\begin{center}\includegraphics[scale=0.45]{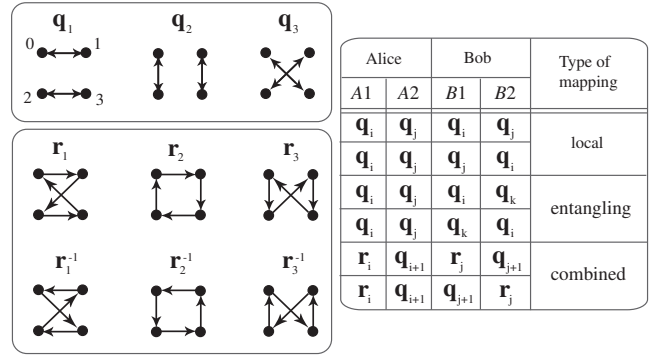}\end{center}
\caption{A schematic representation of the nine possible derangements on four elements represented here by dots. The table indicates whether the state that is mapped out by Bob before the feed forward is applied is locally equivalent to the initial two-qubit state Alice sent, or whether Bob must perform entangling operations on his two-qubit state to reconstruct the transmitted state. This depends on the derangement chosen for each subsystem 1 and 2, and the bus measurement outcomes. Within the two stages of the protocol, the choice of permutations must obey the orthogonality requirements. This explains why we specify the $q_{i+1}$, as it is the only one satisfying the requirements (4), given that $r_i$ was chosen. The table is not exhaustive but gives the main observations.}
\label{fig1}
\end{figure}

In the second part of the protocol, Bob prepares a pair of qubits $B1$ and $B2$ in $|+\rangle=(|0\rangle + |1\rangle)/\sqrt{2}$ states. Upon receiving the bus, he lets the two qubits interact consecutively with it, keeping the same interaction $C=C^{B2}C^{B1}$,
\begin{eqnarray}
C|+\rangle|+\rangle|\xi\rangle=\frac{1}{2}\{|00\rangle_B(x_0|0\rangle+x_1|3\rangle+x_2|1\rangle+x_3|2\rangle) \nonumber \\
+|01\rangle_B(x_0|3\rangle+x_1|0\rangle+x_2|2\rangle+x_3|1\rangle) \nonumber \\
+|10\rangle_B(x_0|1\rangle+x_1|2\rangle+x_2|0\rangle+x_3|3\rangle) \nonumber \\
+|11\rangle_B(x_0|2\rangle+x_1|1\rangle+x_2|3\rangle+x_3|0\rangle)\}.
\label{20}
\end{eqnarray}
To complete the protocol Bob measures the mediating bus in the computational basis. To view the results of different measurement outcomes the above combined state can be written in a matrix form which we term the \textit{pre-measurement matrix}. The pre-measurement matrix contains the possible unitary operations the initial two-qubit state will undergo as it is transmitted in function of the measurement outcomes. Thus defining the projector $\lambda_n=|n\rangle\langle n|$, we rewrite (17) as
\begin{equation}
M_{loc},= \left( \begin{array}{cccc}
\lambda_0 & \lambda_3 & \lambda_1 & \lambda_2\\
\lambda_3 & \lambda_0 & \lambda_2 & \lambda_1\\
\lambda_1 & \lambda_2 & \lambda_0 & \lambda_3\\
\lambda_2 & \lambda_1 & \lambda_3 & \lambda_0\end{array} \right).
\label{21}
\end{equation} 
So for example if Bob measures the bus in the state $|3\rangle$ (corresponding to $\lambda_3$), he has reproduced Alice's initial two-qubit state up to the (known) unitary operation
\begin{equation}
M_{loc,|3\rangle}= \left( \begin{array}{cccc}
0 & 1 & 0 & 0\\
1 & 0 & 0 & 0\\
0 & 0 & 0 & 1\\
0 & 0 & 1 & 0\end{array} \right).
\label{29}
\end{equation} 
Measuring the mediating bus in any one of the states $|0\rangle$, $|1\rangle$, $|2\rangle$, or $|3\rangle$ yields the initial two-qubit state up to the unitary operations $I_{B1}I_{B2}$, $X_{B1}I_{B2}$, $X_{B1}X_{B2}$ and $I_{B1}X_{B2}$ applied to it respectively, where $X$ is the qubit Pauli matrix $X=|1\rangle\langle 0|+|0\rangle\langle 1|$. This means the feed-forward operation $F$ only consists of local unitary operations on the qubits and is therefore a \textit{local mapping}. 

In contrast, if Alice uses the two permutation operators $P^{A1}=Q_1,P^{A2}=Q_3$ and Bob uses $P^{B1}=Q_2,P^{B2}=Q_3$, he then obtains the pre-measurement matrix
\begin{equation}
M_{ent}= \left( \begin{array}{cccc}
\lambda_0 & \lambda_3 & \lambda_1 & \lambda_2\\
\lambda_3 & \lambda_0 & \lambda_2 & \lambda_1\\
\lambda_2 & \lambda_1 & \lambda_3 & \lambda_0\\
\lambda_1 & \lambda_2 & \lambda_0 & \lambda_3\end{array} \right).
\label{21}
\end{equation} 
In this case all measurement outcomes require a non-local feed-forward operation $F$, so we call this an \textit{entangling mapping}. The mapped out state is equivalent to Alice's input state with a CNOT gate applied to it, for all outcomes. 

The second type of interaction makes use of cyclic permutations. We note here that Alice and Bob cannot choose their two permutation operators from the cyclic permutations alone, as they will not fulfill the requirements (4). An example of a valid choice is to set $P^{A1}=P^{B1}=R_1$ and $P^{A2}=P^{B2}=Q_2=R_1^2$, then we obtain the pre-measurement matrix
\begin{equation}
M_{com}= \left( \begin{array}{cccc}
\lambda_0 & \lambda_2 & \lambda_1 & \lambda_3\\
\lambda_2 & \lambda_0 & \lambda_3 & \lambda_1\\
\lambda_1 & \lambda_3 & \lambda_2 & \lambda_0\\
\lambda_3 & \lambda_1 & \lambda_0 & \lambda_2\end{array} \right).
\label{21}
\end{equation} 
Here what we see is that the measurement outcomes $|1\rangle$ and $|3\rangle$ lead to a local mapping while the $|0\rangle$ and $|2\rangle$ measurement outcomes lead to an entangling mapping. For arbitrary states, each mapping occurs with equal probability in this case, we term this measurement dependent case a \textit{combined mapping}. It is worth noting here that either way, the quantum information is left intact, meaning a repeat-until-success scheme \cite{lim} can be envisaged. If the aim of the protocol is to entangle the two transmitted qubits through a CNOT gate, and the permutation operators at hand are those used to generate the output (14), then we can repeat the protocol (on average twice), until the desired entangled output state is obtained.

By searching through different combinations we see that local and entangling mappings can only be achieved if both Alice and Bob choose their permutations from the pairwise $Q_1$, $Q_2$ and $Q_3$ operators. Using the same permutations will yield a local mapping, whereas changing them will yield an entangling mapping. Another important point is that independent of Alice's choice of interaction, Bob using an element $R_i$ will yield a combined mapping. 

As the subsystem dimension increases, finding sets of permutation operators satisfying (4) and observing the feed-forward operations for different measurement outcomes rapidly becomes intractable. Also the entangling power of the resulting unitary operations applied to the transmitted state (before the feed-forward) can vary, unlike in the two-qubit case \cite{clar}. In spite of these difficulties, the general methods given in the next section allow us to systematically investigate higher dimensions. By using criteria characterizing maximally entangling permutation operators \cite{clar}, we obtain sets leading to a maximally entangling mapping in the case of two qutrits (see Appendix).

\section{Building interactions with permutations}

We can generalize the previous discussion, keeping the concepts of local, entangling and combined mappings. To effect these mappings for arbitrary subsystem dimension $d$, we find two different types of interactions based on conditional permutation operators. The first type of interaction makes use of the commuting operators $H$ and $V$ whose corresponding permutations in the cycle notation are
\begin{eqnarray} 
h&=&(0,1,..,d-1)(d,d+1,..,2d-1)\nonumber\\
&&...(d^2-d,d^2-d+1,..,d^2-1),\nonumber\\
&&\nonumber\\
v&=&(0,d,..,d^2-d)(1,d+1,..,d^2-d+2)\nonumber\\
&&...(d-1,2d-1,..,d^2-1),
\label{}
\end{eqnarray}
acting on $d^2$ elements representing the bus basis states. As we can see $h$ and $v$ consist in cycles of length $d$ where each element is included in only one cycle from each. We now identify them with $q_1$ and $q_2$ for $d=2$ respectively. Extending the representation in Fig. 2 we see that if we arrange the elements into a $d\times d$ square lattice, $h$ groups the elements composing the cycles in a horizontal way whereas $v$ groups them in a vertical way. Arbitrary combinations $V^lH^k$ lead to orthogonal permutation operators satisfying (4) and thus we can arrange them into the two sets
\begin{eqnarray}
\{P_k^{B1}|P_k^{B1}=H^k,k=0,..,d-1\},\nonumber \\
\{P_l^{B2}|P_l^{B2}=V^l,l=0,..,d-1\}.
\label{}
\end{eqnarray}
These operators based on permutations with $d$-cycles allow for a transmission of Alice's state without the need for nonlocal operations at the feed-forward stage. This can be seen by first rewriting the bus basis states $|s\rangle$ as $|\mathrm{MOD}_d(s),\left\lfloor s/d\right\rfloor\rangle$ so that the above operators act according to $V^lH^k|m,n\rangle=|\mathrm{MOD}_d(m+k),\mathrm{MOD}_d(n+l)\rangle$. By initiating the bus in the state $|0,0\rangle$, Alice and Bob can choose their sets so that the final state (3) before the bus measurement reads
\begin{eqnarray}
&&\frac{1}{d}\sum_{i,j,k,l=0}^{d-1}x_{ij}\left(|k\rangle|l\rangle\right)_B\left(|i\rangle|j\rangle\right)_A V^{d-l}H^{d-k}V^jH^i |0,0\rangle\nonumber\\
&&=\frac{1}{d}\sum_{i,j,k,l=0}^{d-1}x_{ij}\left(|k\rangle|l\rangle\right)_B\left(|i\rangle|j\rangle\right)_A\nonumber\\ &&\qquad\qquad\qquad\otimes|\mathrm{MOD}_d(i-k),\mathrm{MOD}_d(j-l)\rangle.
\label{}
\end{eqnarray}
Alice measuring her subsystems in the conjugate basis and Bob measuring the bus in the $|m,n\rangle$ state will result in Bobs composite system being in the state
\begin{equation}
\sum_{i,j=0}^{d-1}x_{ij}|\mathrm{MOD}_d(i-m)\rangle|\mathrm{MOD}_d(j-n)\rangle=X^{-m}\otimes X^{-n}|\psi\rangle,
\label{}
\end{equation}
up to local phase corrections induced by Alice's measurements. $|\psi\rangle$ is the initial state of Alice's composite system and $X$ is the generalized Pauli operator \cite{gott1} defined by its action on the basis states: $X|s\rangle\equiv|\mathrm{MOD}_d(s+1)\rangle$. With this interaction we can also choose to deterministically entangle the subsystems in the transmission, directly processing information, as observed in the previous section.

The second type of conditional permutation operator is the simplest and makes use of the cyclic permutation on $d^2$ elements $x=(0,1,...d^2-1)$ corresponding to the generalized Pauli $X$ operator acting on $d^2$ basis states (modulo $d^2$). Because $X^n$ operations commute, the expression (4) becomes a set of simultaneous modulo inequations on different values of $n$. It is always possible to find two sets satisfying these requirements; in the first set, conditioned on the first subsystem we choose 
\begin{equation}
\{P_k^{B1}|P_k^{B1}=X^k,k=0,..,d-1\}.
\label{}
\end{equation}
Based on this choice, we can adapt the second set so that no two combinations induce the same shift operation: 
\begin{equation}
\{P_l^{B2}|P_l^{B2}=X^{ld},l=0,..,d-1\}.
\label{}
\end{equation}
Using this type of permutation again leads to deterministic transfer of Alice's composite system up to a known two-qudit operation. However whether or not Bob's state before the feed-forward is locally equivalent to Alice's input state will depend on the measurement result. We note here that this controlled interaction can be assimilated to the hybrid version of the SUM gate \cite{jam} (acting on qudits of different dimension), the qudit extension of the CNOT gate. 

This cyclic permutation approach can be applied to the generalized case of transmitting $m$ subsystems via a $d^m$-dimensional bus. In this case there are $m$ sets of $d$ permutations (including the identity), each defining the interaction of a particular subsystem with the bus. The main idea behind the expression (4) is conserved: any ordered combination of permutations from the sets (one from each set), must result in a permutation orthogonal to all the other combinations in terms of the Hilbert-Schmidt inner product
\begin{equation}
\mathrm{Tr}\left[\prod_{Bj=1}^{m}P_{k_{Bj}}^{Bj}\left(\prod_{Bj=1}^{m}P_{l_{Bj}}^{Bj}\right)^{\dag}\right]=d^m\prod_{Bj=1}^{m}\delta_{k_{Bj} l_{Bj}},
\label{}
\end{equation}
for all $Bj$, $k$ and $l$, where $Bj$ numbers the subsystems. Following on from the previous case we can use the sets
\begin{eqnarray}
\{P_{k_{Bj}}^j|P_{k_{Bj}}^j=X^{k_{Bj}d^{Bj-1}},Bj=1,..,m, \nonumber\\
\quad \mathrm{and}\quad k_{Bj}=0,..,d-1,\},  
\label{}
\end{eqnarray}
with $X=\sum_{i=0}^{d^m-1}|\mathrm{MOD}_{d^m}(i+1)\rangle\langle i|$, ensuring deterministic state transfer for all $m$ and $d$. The order in which the permutation operators are arranged within the sets will define the feed-forward operation applied by Bob. Thus we have found two types of interactions allowing for the successful transfer of composite systems, with or without entanglement generation. This constitutes a generic resource for quantum data transfer. The physical implementation of the proposed operations along with the coupling between the bus and the subsystems constitute the theme of the next section.

\section{A continuous variable bus}
The implementation of general qudit gates requires considerable control. However the second interaction with which we propose to implement our protocols only depends on the ability to perform a generalized $X$ operation conditionally. The use of a continuous variable bus may seem like a complication at first, but interestingly it provides a very natural way of realizing such a conditional operation. Given an interaction Hamiltonian of the form $H_{int}=-\hbar\chi \hat{N}_{bus}\hat{\Lambda}_{sub}$ where $\hat{\Lambda}_{sub}=\sum_{s=0}^{d-1}s|s\rangle\langle s|$ acts on the subsystem, we can approximate the conditional $X$ by preparing the bus in a coherent state $|\alpha\rangle$. $\hat{N}_{bus}$ represents the number operator acting on the energy eigenstates of the harmonic oscillator as $\hat{N}_{bus}|n\rangle=n|n\rangle$. After an interaction time $t$ the combined state evolves as $e^{-iH_{int}t/\hbar}|s\rangle|\alpha\rangle=|s\rangle|\alpha e^{i\theta s}\rangle$ with $\theta=\chi t$. Thus we see that the possible states of the subsystem are encoded into the phase of the coherent state. 

Now given a $D$-dimensional composite system, the bus states we will write as $\{|n)\equiv|\alpha e^{i2n\pi/D}\rangle,n=0,..,D-1\}$. It is worth noting here that this set of states is literally generated at the encoding stage, on Alice's side in the transfer protocol or on Bob's side in the teleportation protocol. Each subsystem interacts with the bus for a different amount of time, rotating the states of the bus in phase space by a different angle (see Fig. 3(a)). Setting $\theta=2\pi/D$, we can view the effect of a general interaction as

\begin{equation}
C^k|s\rangle|n)=|s\rangle|\mathrm{MOD}_D(n+ks)),
\label{}
\end{equation}
with $C=e^{i\theta\hat{N}_{bus}\hat{\Lambda}_{sub}}$. Thus we see that by repeating the interactions or equivalently increasing the interaction time, we obtain all the conditional operations required to implement our protocol, even for arbitrary numbers of subsystems (22). This is achieved through the cyclic nature of the rotation operation $e^{i\theta \hat{N}}$ on the coherent state.

\begin{figure}[!htb]
\begin{center}\includegraphics[scale=0.4]{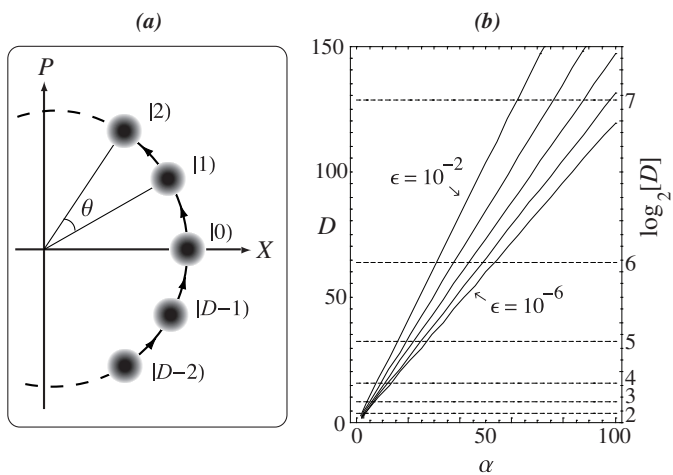}\end{center}
\caption{(a) A phase space picture of the cyclic effect of the shift operation $e^{i\theta \hat{N}}$ on the state of the continuous variable bus, with $\theta=2\pi/D$. As defined in the text, $|n)=|\alpha e^{in\theta}\rangle$. (b) The maximum dimension of the composite system to be transmitted as a function of the amplitude $\alpha$, for a fixed overlap $\epsilon=10^{-j}$ representing the error. From top to bottom we have $j=2,..,6$. The dashed horizontal lines represent the capacity of the channel in number of qubits that can be transferred.}
\label{fig1}
\end{figure}

The bus states do not form an orthogonal basis $(n|m)\neq \delta_{nm}$, and so the dimension $D$ of the transmitted composite system will be limited by the available amplitude $\alpha$ of the bus. For a fixed overlap $\epsilon=(n|n+1)$ which is deemed acceptable, the dimension of the composite system is bounded from above by
\begin{equation}
D\leq\frac{2\pi}{\mathrm{cos}^{-1}(\mathrm{ln}\epsilon/\alpha^2+1)}.
\label{}
\end{equation}
The behavior of this bound is illustrated in Fig. 3(b). We can see that the scaling is close to being linear and the capacity of the bus is large, even for $\epsilon$ as low as $10^{-5}$. In this case the continuous variable bus can potentially teleport up to 7 qubits with a moderate amplitude of $\alpha=100$.

\section{conclusion} 

In this work we proposed and examined the use of a passive mediating bus for the transmission of quantum information over composite systems. We found conditional permutation operations allowing a deterministic transfer of information for subsystems of arbitrary dimension. With the first type of permutations we can choose to keep the transmitted composite system in a locally equivalent form (local mapping), minimizing the work required at the feed-forward stage. On the other hand we can choose to entangle the subsystems in the transmission (entangling mapping), using the bus to implement generalized teleportation based quantum computation. The second type of interaction allowing for the transfer of arbitrary numbers of subsystems relies on cyclic permutations (associated with the generalized $X$ Pauli operator), reminiscent of the hybrid qudit SUM gate. We investigated both of these interactions in the two-qubit and two-qutrit cases. Finally we observed how a continuous variable bus could be used to implement our protocols through a simple interaction.

The protocols lead to the physical compression of quantum information, enabling efficient and flexible data transfer. This is a clear asset for the realization of a large scale quantum computer. Future work will involve adapting the protocols for the transmission of hybrid composite systems (subsystems of different dimensions) and a more detailed analysis of possible physical implementations.

\section{Acknowledgements}
SGRL acknowledges the support of a Monbugakusho scholarship. ADG acknowledges a JSPS Invitation Fellowship, is the recipient of an Australian Research Council Queen Elizabeth II Fellowship (DP0880466) and is in part supported by the Australian Government, the U.S. National Security Agency (NSA), Advanced Research and Development Activity (ARDA), Army Research Office (ARO) under Contracts No. W911NF-04-1-0290. This work was supported in part by MEXT in Japan and the EU project QAP. 
\\
\begin{center}
    {\bf APPENDIX: TRANSMITTING TWO QUTRITS}
  \end{center}
  
Alice holds two qutrits initially unentangled with the bus, with basis states $\left\{|0\rangle,|1\rangle,|2\rangle\right\}$ for each qutrit and $\left\{|s\rangle,d=0,1,..,8\right\}$ for the bus. The three systems are coupled via the consecutive interactions
\begin{eqnarray}
\hat{C}=\left(|0\rangle\langle0|\otimes I^{A2}+|1\rangle\langle1|\otimes P_1^{A2}+|2\rangle\langle2|\otimes P_2^{A2}\right)\nonumber\\
\times\left(|0\rangle\langle0|\otimes I^{A1}+|1\rangle\langle1|\otimes P_1^{A1}+|2\rangle\langle2| \otimes P_2^{A1}\right),
\label{}
\end{eqnarray}
where qutrit $A1$ interacts with the bus before qutrit $A2$. We now must find sets of permutations $\left\{I,P_1,P_2\right\}^{A1}$ and $\left\{I,P_1,P_2\right\}^{A2}$ which satisfy the requirements for complete information transfer (4).

Following the first type of interaction proposed in section III, we identify nine orthogonal permutation operators including the identity. The two operators $G$ and $H$ generating all nine of them when combined, correspond to the derangements
\begin{equation}
g=(012)(345)(678)\;\;\mathrm{and}\;\;h=(036)(147)(258).
\label{}
\end{equation} 
We write the permutation operators as $Y_{n,m}=H^nG^m$ with $n,m=0,1,2$. By combining them correctly we can satisfy the relations (4) and thus realize a deterministic transfer of the two-qutrit state. We illustrate this with a first example, in which Alice couples her input composite system to the bus via the operators $\{I,Y_{0,1},Y_{0,2}\}^{A1}$ and $\{I,Y_{1,0},Y_{2,0}\}^{A2}$. After Alice measures out her two qutrits the bus is in the state
\begin{eqnarray}
|\xi\rangle=x_0|0\rangle+x_1|3\rangle+x_2|6\rangle+x_3|1\rangle+x_4|4\rangle\nonumber\\
+x_5|7\rangle+x_6|2\rangle+x_7|5\rangle+x_8|8\rangle
\label{}
\end{eqnarray} 
up to phase corrections. Bob then prepares two blank qutrits each in the superposition $(|0\rangle+|1\rangle+|2\rangle)/\sqrt{3}$ and couples them to the bus via the inverse permutations $\{I,Y_{0,2},Y_{0,1}\}^{B1}$ and $\{I,Y_{2,0},Y_{1,0}\}^{B2}$. This yields the pre-measurement matrix
\begin{equation}
M= \left( \begin{array}{ccc|ccc|ccc}
\lambda_0 & \lambda_3 & \lambda_6 & \lambda_1 & \lambda_4 & \lambda_7 & \lambda_2 & \lambda_5 & \lambda_8\\
\lambda_6 & \lambda_0 & \lambda_3 & \lambda_7 & \lambda_1 & \lambda_4 & \lambda_8 & \lambda_2 & \lambda_5\\
\lambda_3 & \lambda_6 & \lambda_0 & \lambda_4 & \lambda_7 & \lambda_1 & \lambda_5 & \lambda_8 & \lambda_2\\ \hline
\lambda_2 & \lambda_5 & \lambda_8 & \lambda_0 & \lambda_3 & \lambda_6 & \lambda_1 & \lambda_4 & \lambda_7\\
\lambda_8 & \lambda_2 & \lambda_5 & \lambda_6 & \lambda_0 & \lambda_3 & \lambda_7 & \lambda_1 & \lambda_4\\
\lambda_5 & \lambda_8 & \lambda_2 & \lambda_3 & \lambda_6 & \lambda_0 & \lambda_4 & \lambda_7 & \lambda_1\\ \hline
\lambda_1 & \lambda_4 & \lambda_7 & \lambda_2 & \lambda_5 & \lambda_8 & \lambda_0 & \lambda_3 & \lambda_6\\
\lambda_7 & \lambda_1 & \lambda_4 & \lambda_8 & \lambda_2 & \lambda_5 & \lambda_6 & \lambda_0 & \lambda_3\\
\lambda_4 & \lambda_7 & \lambda_1 & \lambda_5 & \lambda_8 & \lambda_2 & \lambda_3 & \lambda_6 & \lambda_0\end{array} \right).
\label{21}
\end{equation} 
This is a local mapping, i.e. Bob obtained Alice's two-qutrit input state up to local operations, independent of the measurement outcome. By using this set of permutation operators, we can also achieve an entangling mapping. Starting with the same interactions on Alice's side but switching to $\{Y_{0,1},Y_{0,2}\}^{B1}$ and $\{I,Y_{2,2},Y_{1,1}\}^{B2}$ on Bob's side we obtain the pre-measurement matrix
\begin{equation}
M= \left( \begin{array}{ccc|ccc|ccc}
\lambda_0 & \lambda_3 & \lambda_6 & \lambda_1 & \lambda_4 & \lambda_7 & \lambda_2 & \lambda_5 & \lambda_8\\
\lambda_8 & \lambda_2 & \lambda_5 & \lambda_6 & \lambda_0 & \lambda_3 & \lambda_7 & \lambda_1 & \lambda_4\\
\lambda_4 & \lambda_7 & \lambda_1 & \lambda_5 & \lambda_8 & \lambda_2 & \lambda_3 & \lambda_6 & \lambda_0\\ \hline
\lambda_1 & \lambda_4 & \lambda_7 & \lambda_2 & \lambda_5 & \lambda_8 & \lambda_0 & \lambda_3 & \lambda_6\\
\lambda_6 & \lambda_0 & \lambda_3 & \lambda_7 & \lambda_1 & \lambda_4 & \lambda_8 & \lambda_2 & \lambda_5\\
\lambda_5 & \lambda_8 & \lambda_2 & \lambda_3 & \lambda_6 & \lambda_0 & \lambda_4 & \lambda_7 & \lambda_1\\ \hline
\lambda_2 & \lambda_5 & \lambda_8 & \lambda_0 & \lambda_3 & \lambda_6 & \lambda_1 & \lambda_4 & \lambda_7\\
\lambda_7 & \lambda_1 & \lambda_4 & \lambda_8 & \lambda_2 & \lambda_5 & \lambda_6 & \lambda_0 & \lambda_3\\
\lambda_3 & \lambda_6 & \lambda_0 & \lambda_4 & \lambda_7 & \lambda_1 & \lambda_5 & \lambda_8 & \lambda_2\end{array} \right).
\label{21}
\end{equation} 
Each measurement outcome will simulate an entangling operation on the transmitted state. Clarisse \textsl{et al.} \cite{clar} derived criteria for identifying maximally entangling permutation matrices (acting on two systems of equal dimension), which we review here. The matrix corresponding to a permutation operator $P$ is maximally entangling over all unitary operations if it satisfies the following conditions: every block contains a single nonzero entry; all blocks are different; nonzero entries in the same block-row are in different subcolumns; nonzero entries in the same block-column are in different subrows. In the case of two qubits, the CNOT operation constitutes a maximally entangling permutation. 

From these criteria it can be seen that the above resulting matrix is \textit{not} maximally entangling (for all measurement outcomes), because it fails to fulfill one of the requirements: one identifies identical blocks. However with a judicious choice of permutations, one can achieve a maximally entangling mapping. For example Alice choosing the sets $\left\{I,Y_{0,1},Y_{0,2}\right\}^{A1}$, $\left\{I,Y_{1,0},Y_{2,0}\right\}^{A2}$, and Bob the sets $\left\{I,Y_{2,1},Y_{1,2}\right\}^{B1}$ and $\left\{I,Y_{2,2},Y_{1,1}\right\}^{B2}$ results in the pre-measurement matrix
\begin{equation}
M_{max}= \left( \begin{array}{ccc|ccc|ccc}
\lambda_0 & \lambda_3 & \lambda_6 & \lambda_1 & \lambda_4 & \lambda_7 & \lambda_2 & \lambda_5 & \lambda_8\\
\lambda_8 & \lambda_2 & \lambda_5 & \lambda_6 & \lambda_0 & \lambda_3 & \lambda_7 & \lambda_1 & \lambda_4\\
\lambda_4 & \lambda_7 & \lambda_3 & \lambda_5 & \lambda_8 & \lambda_2 & \lambda_3 & \lambda_6 & \lambda_0\\ \hline
\lambda_7 & \lambda_1 & \lambda_4 & \lambda_8 & \lambda_2 & \lambda_5 & \lambda_6 & \lambda_0 & \lambda_3\\
\lambda_3 & \lambda_6 & \lambda_0 & \lambda_4 & \lambda_7 & \lambda_1 & \lambda_5 & \lambda_8 & \lambda_2\\
\lambda_2 & \lambda_5 & \lambda_8 & \lambda_0 & \lambda_3 & \lambda_6 & \lambda_1 & \lambda_4 & \lambda_7\\ \hline
\lambda_5 & \lambda_8 & \lambda_2 & \lambda_3 & \lambda_6 & \lambda_0 & \lambda_4 & \lambda_7 & \lambda_1\\
\lambda_1 & \lambda_4 & \lambda_7 & \lambda_2 & \lambda_5 & \lambda_8 & \lambda_0 & \lambda_3 & \lambda_6\\
\lambda_6 & \lambda_0 & \lambda_3 & \lambda_7 & \lambda_1 & \lambda_4 & \lambda_8 & \lambda_2 & \lambda_5\end{array} \right).
\label{21}
\end{equation} 
Here all blocks are different and for each measurement outcome we have a maximally entangling permutation operator and in consequence a maximally entangling unitary \cite{clar}, acting on the transmitted qutrits.

Continuing with the second method of section III we now use the shift operation $X=\sum_{n=0}^{8}|n+m\;(\mathrm{mod}\;9)\rangle\langle n|$, the sets are of the form $\left\{I,X,X^2\right\}^{A1}$ and $\left\{I,X^3,X^6\right\}^{A2}$. If Alice uses the ordered combination above and Bob couples his two qutrits to the bus with the combination $\left\{I,X^8,X^7\right\}^{B1}$ and $\left\{I,X^6,X^3\right\}^{B2}$ (i.e. the inverse, which is also a solution to (4)) we obtain the pre-measurement matrix
\begin{equation}
M= \left( \begin{array}{ccc|ccc|ccc}
\lambda_0 & \lambda_3 & \lambda_6 & \lambda_1 & \lambda_4 & \lambda_7 & \lambda_2 & \lambda_5 & \lambda_8\\
\lambda_6 & \lambda_0 & \lambda_3 & \lambda_7 & \lambda_1 & \lambda_4 & \lambda_8 & \lambda_2 & \lambda_5\\
\lambda_3 & \lambda_6 & \lambda_0 & \lambda_4 & \lambda_7 & \lambda_1 & \lambda_5 & \lambda_8 & \lambda_2\\ \hline
\lambda_8 & \lambda_2 & \lambda_5 & \lambda_0 & \lambda_3 & \lambda_6 & \lambda_1 & \lambda_4 & \lambda_7\\
\lambda_5 & \lambda_8 & \lambda_2 & \lambda_6 & \lambda_0 & \lambda_3 & \lambda_7 & \lambda_1 & \lambda_4\\
\lambda_2 & \lambda_5 & \lambda_8 & \lambda_3 & \lambda_6 & \lambda_0 & \lambda_4 & \lambda_7 & \lambda_1\\ \hline
\lambda_7 & \lambda_1 & \lambda_4 & \lambda_8 & \lambda_2 & \lambda_5 & \lambda_0 & \lambda_3 & \lambda_6\\
\lambda_4 & \lambda_7 & \lambda_1 & \lambda_5 & \lambda_8 & \lambda_2 & \lambda_6 & \lambda_0 & \lambda_3\\
\lambda_1 & \lambda_4 & \lambda_7 & \lambda_2 & \lambda_5 & \lambda_8 & \lambda_3 & \lambda_6 & \lambda_0\end{array} \right).
\label{21}
\end{equation} 
The same observation as in the two qubit case can be made. Different measurement outcomes call for different types of feed forward. If we measure the states $|0\rangle$, $|3\rangle$ or $|6\rangle$ (which occurs with a probability 1/3) we obtain the initial state up to local operations on the two qutrits. However all other outcomes will lead to the initial state having undergone an entangling operation, though not a maximally entangling one.
\\
\\
\\


\begin{thebibliography}{99}
\bibitem{niel-chuang} M.A. Nielsen and I.L. Chuang, Quantum Computation and Quantum Information, Cambridge University Press (2000).
\bibitem{shor}
P.W. Shor, Proc. 35th Annual Symposium on the Foundations of Computer Science, p.124 (1994).
\bibitem{ekert}
A. Ekert and R. Jozsa, Rev. Mod. Phys. \textbf{68}, 733 (1996).
\bibitem{grover}
L. Grover, Proc. 28th Annual ACM Symposium on the Theory of Computing, p. 212-219 (1996).
\bibitem{hol} 
L.C.L. Hollenberg, A.D. Greentree, A.G. Fowler, and C.J. Wellard, Phys. Rev. B \textbf{74}, 045311 (2006).
\bibitem{thaker} 
D.D. Thaker, T.S. Metodi, A.W. Cross, I.L. Chuang, and F.T. Chong, ACM Proc. 33rd Annual International Symposium on Computer Architecture (2006).
\bibitem{rod}
R. Van Meter and M. Oskin, ACM Journal on Emerging Technologies in Computing Systems, vol. \textbf{2} Issue 1 (2006).
\bibitem{kielp}
D. Kielpinksy, C. Monroe, and D.J. Wineland, Nature (London) \textbf{417}, 709 (2002).
\bibitem{tayl}
J.M. Taylor, H.-A. Engel, W. Dur, A. Yacoby, C.M. Marcus, P. Zoller, and M.D. Lukin, Nat. Phys. \textbf{1}, 177 (2005).
\bibitem{div}
D.P. DiVincenzo, Fortschr. Phys. \textbf{48} 771 (2000).
\bibitem{skinner}
A.J. Skinner, M.E. Davenport, and B.E. Kane, Phys. Rev. Lett. \textbf{90}, 087901 (2003).
\bibitem{duan1}
L.-M. Duan, J.I. Cirac, P. Zoller, and E.S. Polzik, Phys. Rev. Lett. \textbf{85}, 5643 (2000).
\bibitem{duan2}
L.-M. Duan, and H.J. Kimble, Phys. Rev. Lett. \textbf{90}, 253601 (2003).
\bibitem{oi}
D.K.L. Oi, S.J. Devitt, and L.C.L. Hollenberg, Phys. Rev. A. \textbf{74}, 052313 (2006).
\bibitem{bose} S. Bose, Phys. Rev. Lett. \textbf{91}, 207901 (2003).
\bibitem{lloyd}
S. Lloyd, Phys. Rev. Lett. \textbf{90}, 167902 (2003).
\bibitem{fries}
M. Friesen, A. Biswas, X. Hu, and D. Lidar, Phys. Rev. Lett. \textbf{98}, 230503 (2007).
\bibitem{tim}
T.P. Spiller, K. Nemoto, S.L. Braunstein, W.J. Munro, P. van Loock and G.J. Milburn, New J. Phys. \textbf{8}, 30 (2006).
\bibitem{loock-rep} 
P. van Loock, T.D. Ladd, K. Sanaka, F. Yamaguchi, K. Nemoto, W.J. Munro, Y. Yamamoto, Phys. Rev. Lett. \textbf{96}, 240501 (2006).
\bibitem{gotc}
D. Gottesman and I.L. Chuang, Nature \textbf{402}, 390 (1999).
\bibitem{rod-multi}
R. Van Meter, K. Nemoto, and W.J. Munro, IEEE Transactions on Computers \textbf{56}(12), 1643 (2007).

\bibitem{andy}
A.D. Greentree, S.G. Schirmer, F. Green, L.C.L. Hollenberg, A.R. Hamilton, and R.G. Clark, Phys. Rev. Lett. \textbf{92}, 097901 (2004).
\bibitem{dev}
S.J. Devitt, S.G. Schirmer, D.K.L. Oi, J.H. Cole, and L.C.L. Hollenberg, New J. Phys. \textbf{9}, 384 (2007).
\bibitem{bart}
S.D. Bartlett, H. de Guis, and B.C. Sanders, Phys. Rev A \textbf{65}, 052316 (2002).
\bibitem{jam}
J. Daboul, X. Wang and B.C. Sanders, J. Phys. A: Math. Gen. \textbf{36} (14), 2525-2536 (2003).
\bibitem{leary}
D.P. O'Leary, G.K. Brennen and S.S. Bullock, Phys. Rev. A \textbf{74}, 032334 (2006).
\bibitem{xen} 
X.-H. Li, F.-G. Deng and H.-Y. Zhou, Chinese Physics Letters \textbf{24}, 1151 (2007).
\bibitem{hugh}
D.Mc Hugh and J. Twamley, New J. Phys. \textbf{7}, 174 (2005).
\bibitem{ralph}
T.C. Ralph, K.J. Resch, and A. Gilchrist, Phys. Rev. A \textbf{75}, 022313 (2007).
\bibitem{bauer}
P.H.E. Meijer, {\it Group Theory, The Application to Quantum Mechanics} (North-Holland Publishing, 1962).
\bibitem{mont}
P. R. de Montmort, Essai d'analyse sur les jeux de hasard. Paris (1708).
\bibitem{lim} Y.L. Lim, A. Beige, L.C. Kwek, Phys. Rev. Lett. \textbf{95}, 030505 (2005).
\bibitem{clar}
L. Clarisse, S. Ghosh, S. Severini and Anthony Sudbery, Phys. Rev. A \textbf{72}, 012314 (2005).
\bibitem{gott1}
D. Gottesman, Chaos, Solitons and Fractals \textbf{10} 1749 (1999).

\end{thebibliography}
\end{document}